\documentclass[10pt,preprint]{emulateapj}
\usepackage{url}
\usepackage{rotating}
\usepackage{color}
\usepackage{natbib}
\newcommand{\fig}[1]{Figure~\ref{#1}}
\newcommand{\tbl}[1]{Table~\ref{#1}}
\newcommand{\speed}[1]{#1 km~s${}^{-1}$}

\newcommand{\tabincell}[2]{\begin{tabular}{@{}#1@{}}#2\end{tabular}}

\begin{document}

\shorttitle{} %

\shortauthors{Shen et al.}

\title{Simultaneous Transverse Oscillations of a Prominence and a Filament and Longitudinal Oscillation of another Filament Induced by a Single Shock Wave}

\author{Yuandeng Shen\altaffilmark{1,2,3,4}, Ying D. Liu\altaffilmark{2}, P. F. Chen\altaffilmark{4}, and Kiyoshi Ichimoto\altaffilmark{3}}

\altaffiltext{1}{Yunnan Observatories, Chinese Academy of Sciences, Kunming 650011, China; ydshen@ynao.ac.cn}
\altaffiltext{2}{State Key Laboratory of Space Weather, Chinese Academy of Sciences, Beijing 100190, China}
\altaffiltext{3}{Kwasan and Hida Observatories, Kyoto University, Yamashina-ku, Kyoto 607-8471, Japan}
\altaffiltext{4}{Key Laboratory of Modern Astronomy and Astrophysics, School of Astronomy \& Space Science, Nanjing University, Nanjing 210093, China}

\begin{abstract}
We present the first stereoscopic and Doppler observations of simultaneous transverse oscillations of a prominence and a filament and longitudinal oscillation of another filament launched by a single shock wave. Using H$\alpha$ Doppler observations, we derive the three-dimensional oscillation velocities at different heights along the prominence axis. The results indicate that the prominence has a larger oscillation amplitude and damping time at higher altitude, but the periods at different heights are the same (i.e., 13.5 minutes). This suggests that the prominence oscillates like a linear vertical rigid body with one end anchored on the Sun. One of the filaments shows weak transverse oscillation after the passing of the shock, which is possibly due to the low altitude of the filament and the weakening (due to reflection) of the shock wave before the interaction. Large amplitude longitudinal oscillation is observed in the other filament after the passing of the shock wave. The velocity amplitude and period are about \speed{26.8} and 80.3 minutes, respectively. We propose that the orientation of a filament or prominence relative to the normal vector of the incoming shock should be an important factor for launching transverse or longitudinal filament oscillations. In addition, the restoring forces of the transverse prominence are most likely due to the coupling of gravity and magnetic tension of the supporting magnetic field, while that for the longitudinal filament oscillation is probably the resultant force of gravity and magnetic pressure.
\end{abstract}


\section{Introduction}
Solar prominences or filaments are cool and dense plasmas supported by coronal magnetic fields, and their eruption often causes large-scale reconfiguration of coronal magnetic fields and is usually the source of solar storms that can influence the near-Earth space environmental condition \citep[e.g.,][]{chen11,fang12,shenc12,wang13,liuy14}. For the past several decades, the dynamics of solar prominences has been the subject of a large number of studies. However, many fundamental problems about prominences are still open questions \citep[][]{shen11b,shen12d,shen12e,bi12,bi13a,pare14}. Therefore, the investigation of the basic physics in prominence is of particular importance in solar physics. In the present paper, we focus on the intriguing large amplitude oscillatory motion in prominences/filaments, which is rarely observed, and many questions about this phenomenon are yet to be understood \citep{trip09}. On the other hand, many studies have indicated that small amplitude oscillation in prominences is a common phenomenon, which is thought to be locally generated by perturbations omnipresent in the photosphere and chromosphere \citep{arre12}. Prominences and filaments are the same entities with the former seen in emission over the solar limb and the latter on the disk in absorption. We will use the two terms interchangeably throughout the paper.

Large amplitude oscillatory motion in prominences has been observed for about eighty years since the early reports in the 1930s \citep{dyso30,newt35}. Previous observations have indicated that oscillating filaments are tightly in association with remote flare activities and global shock waves \citep{more60,shen12b,shen12c,shen13a,yang13,xue13,liuw14}. Due to the interaction with shock waves, a filament usually starts to oscillate with a large downward motion and then experiences an up-and-down damping harmonic oscillation process \citep{dods49,bruz51,more60,shen14,xue14}. If one observes an oscillating filament using H$\alpha$ line-wings, the filament will periodically appear and disappear in the line-wing due to the Doppler effect. Therefore, such kind of oscillating filament is also called ``winking filaments'' in history. \cite{rams66} studied 11 winking filaments and found that filaments always oscillate with their own characteristic frequency, which is determined by the inherent property of the filament rather than the disturbance agents. Previous studies have indicated that winking filaments are usually initiated by fast magnetohydrodynamics (MHD) shocks in the solar atmosphere, for example, chromosphere Moreton waves \citep{more60} and coronal EUV waves \citep{shen12b,shen14}. Since the speed of a fast MHD wave is faster in the corona than that in the chromosphere, the normal vector of the wavefront points downward as the wave travels from the flare site \citep{uchi68,liu12,liu13}. Therefore, the arriving of a shock wave will push the filament downwards to a certain displacement at the first, and then it will oscillate for a few cycles with the coupling of gravity and magnetic forces as the restoring force. According to the classification of large amplitude filament oscillations, the winking filaments belong to transverse and/or vertical filament oscillation, whose oscillation direction is perpendicular to the filament's main axis. In many studies, the chromosphere Moreton wave is often thought as the trigger of winking filaments \citep[][]{eto02}. However, coronal EUV waves could also trigger the oscillations of some filaments \citep{okam04,kuma13,shen14}, and coronal loops \citep{shen12b,kuma13,zhen13}. Very recently, \cite{shen14} observed a chain of winking filaments in association with a remote flare and a bright EUV wave in the west hemisphere. In this event, there is no chromosphere Moreton wave that could be detected in H$\alpha$ observations, and the EUV wave is also difficult to identify at the location of a long filament in the east hemisphere. The authors found that the initiation of the filament oscillation is in agreement with the inferred arriving time of the lateral surface EUV wave, which is driven by the associated coronal mass ejection (CME) at the first and then freely propagates in the corona along the solar surface. This observations suggests that some winking filaments can be triggered by invisible (weak) shock waves. According the table in \cite{trip09} and \cite{shen14}, the velocity amplitude, period, and damping time of winking filaments are in ranges of \speed{6 -- 41}, 11 -- 29, and 25 -- 180 minutes, respectively. Theoretical investigations of transverse/vertical filament oscillations have also been performed by \cite{hyde66} and \cite{klec69}, in which the magnetic tension is considered as the main restoring force, while the damping can be attributed either to energy losses by emission of shock waves into the ambient corona or to various dissipative processes \citep[][and the references therein]{trip09}.

Besides large amplitude transverse filament oscillations, \cite{jing03} first reported the other type of large amplitude oscillations in filaments, i.e., longitudinal filament oscillation, where the oscillatory motion is along the filament axis. According to previous studies, the velocity amplitude, period, and damping time of longitudinal oscillations are in ranges of \speed{30 -- 100}, 44 -- 160, and 115 -- 600 minutes, respectively. Previous observations have indicated that the initiation of longitudinal oscillation in filaments is often related to the nearby micro flare or jet activities \citep{jing03,jing06,vrsn07,li12,bi14}. However, what is the main restoring force is still an open question. Generally speaking, there are three candidate forces that could be the restoring force for longitudinal filament oscillations, i.e., gravity, magnetic pressure, and magnetic tension forces. \cite{vrsn07} explained the trigger of longitudinal filament oscillations in terms of poloidal magnetic flux injection into the filament by magnetic reconnection at one of the filament legs, and the magnetic pressure gradient along the filament axis is proposed to be the restoring force. However, \cite{li12} suggested that the restoring force is probably the coupling of the magnetic tension and gravity. Theoretical works have also been done for investigating the restoring force and damping mechanisms of large amplitude longitudinal filament oscillations \citep[][]{luna12a,luna12b,luna14,zhan12,zhan13}. The authors found that the main restoring force is mainly the projection component of the gravity along the filament axis, while the damping is possibly related to the wave leakage, mass accretion, and geometry of the magnetic structure which supports the filament mass. It should be noted that in these numerical studies the authors did not take the filament magnetic field into account.

In addition to the above mentioned large amplitude transverse and longitudinal filament oscillations, large amplitude long and ultra-long-period filament oscillations are also observed (say, 8 -- 27 hours), which is interpreted in terms of slow string MHD modes and/or thermal over-stability associated with peculiarities of the cooling/heating function \citep{foul04,foul09}. On October 15, 2002, the oscillation of a polar crown filament was observed before and during the slow-rising pre-eruption phase, and the period is about 2.5 hours \citep{isob06,isob07,pint08,chen08}. The authors proposed that the fast magnetic reconnection that changes the equilibrium of the supporting magnetic system could be the possible trigger mechanism. These observations of prominence oscillations can provide a powerful diagnostic tool for the forecasting of prominence eruption. The observation of large amplitude prominence oscillations is important for studying the property of prominences and the surrounding coronal plasma with the so-called prominence seismology \citep[e.g.,][]{hyde66,uchi70,isob07,pint08,hers11}. In addition, it is also important for diagnosing the characteristics and the arriving of large-scale MHD waves in the solar atmosphere \citep[e.g.,][]{eto02,okam04,gilb08,gosa12,jack13,shen14}. For more background knowledge on filament oscillations, we refer to several recent reviews \citep{oliv02,trip09,arre12}.

In the present paper, we present the first observations of simultaneous transverse oscillations  of a prominence and a filament and longitudinal oscillation of another filament launched by the  shock wave in association with the impulsive {\it GOES} X6.9 flare on August 09, 2011. This flare is so far the most powerful one in the current solar cycle 24, and it has attracted a lot of attention of solar physicists. \cite{shen12c} and \cite{asai12} studied the associated shock wave in different layers of the solar atmosphere, in which the authors firstly reported the corresponding wave signature at the bottom layer of the chromosphere, which suggests that a stronger shock wave can reach down deeper to the solar atmosphere. This event also caused the oscillation of a nearby loop system \citep{sriv13}, the peculiar phenomenon of microwave quasi-periodic pulsation \citep{tan12}, and influence on the Earth's electrodynamics of the equatorial and low-latitude ionosphere \citep{srip13}. Although \cite{asai12} have briefly studied the associated prominence oscillation in this event, here, we present a more detailed analysis about the prominence and filament oscillations involved in this event, with Doppler and stereoscopic observations taken by the Solar Magnetic Activity Research Telescope \citep[SMART;][]{ueno04}, the {\em Solar Dynamics Observatory} \citep[{\em SDO};][]{pesn12}, and the {\em Solar TErrestrial RElations Observatory} \citep[{\em STEREO};][]{kais08}. Instruments and data sets are described in Section 2, results are presented in Section 3, and discussions and conclusions are given in Section 4.

\section{Instruments and Data Sets}
In this paper, we use stereoscopic and Doppler multi-wavelength observations to study the simultaneous prominence and filament oscillations caused by the global shock wave in association with the X6.9 flare on August 09, 2012. The used instruments include SMART,  {\em SDO}, and {\em STEREO}-Ahead, and the flare light-curves recorded by the {\em Geostationary Operational Environmental Satellite} ({\em GOES}) and the {\em Reuven Ramaty High Energy Solar Spectroscopic Imager} \citep[{\em RHESSI};][]{lin02}. The SMART is a ground-based telescope at Hida Observatory of Kyoto University, Japan, which is designed to observe the dynamic chromosphere and the photospheric magnetic field. We use the full-disk H$\alpha$ images taken in seven wavelengths at H$\alpha$ center, $\pm 0.5$ \AA, $\pm 0.8$ \AA, and $\pm 1.2$ \AA. The telescope takes images with the seven channels near simultaneously ($< 1$ min), and the field-of-view (FOV) and pixel resolution are $2300\arcsec \times 2300\arcsec$ and $0\arcsec.6$, respectively. The cadence of the images is 2 minutes. The high temporal and spatial resolution EUV observations are taken by the Atmospheric Imaging Assembly \citep[AIA;][]{leme12} onboard {\em SDO}, which provides continuous images at seven EUV and three UV-visible channels, and the cadence and pixel size are 12 seconds and $0\arcsec.6$, respectively. The full-disk EUV observations taken by {\em STEREO}-Ahead are also used in this paper. On August 09, 2011,  the separation angle between {\em STEREO}-Ahead and the Earth is about $101^{\circ}$. Since the source region located close to the west limb from the Earth's view angle, the event was not observed by the {\em STEREO}-Behind. The pixel size of {\em STEREO} images is $1\arcsec.58$, while the cadences of the 195 \AA\, 304 \AA\, and 171 \AA\ images are 5, 10, and 120 minutes, respectively. All images used in this paper are differentially rotated to a reference time of 08:05 UT by using the standard SolarSoftWare (SSW), and the north is up and east is to the left.

\section{Results}
\subsection{Overview of the Event}
\begin{figure}[thbp]
\epsscale{0.9}
\plotone{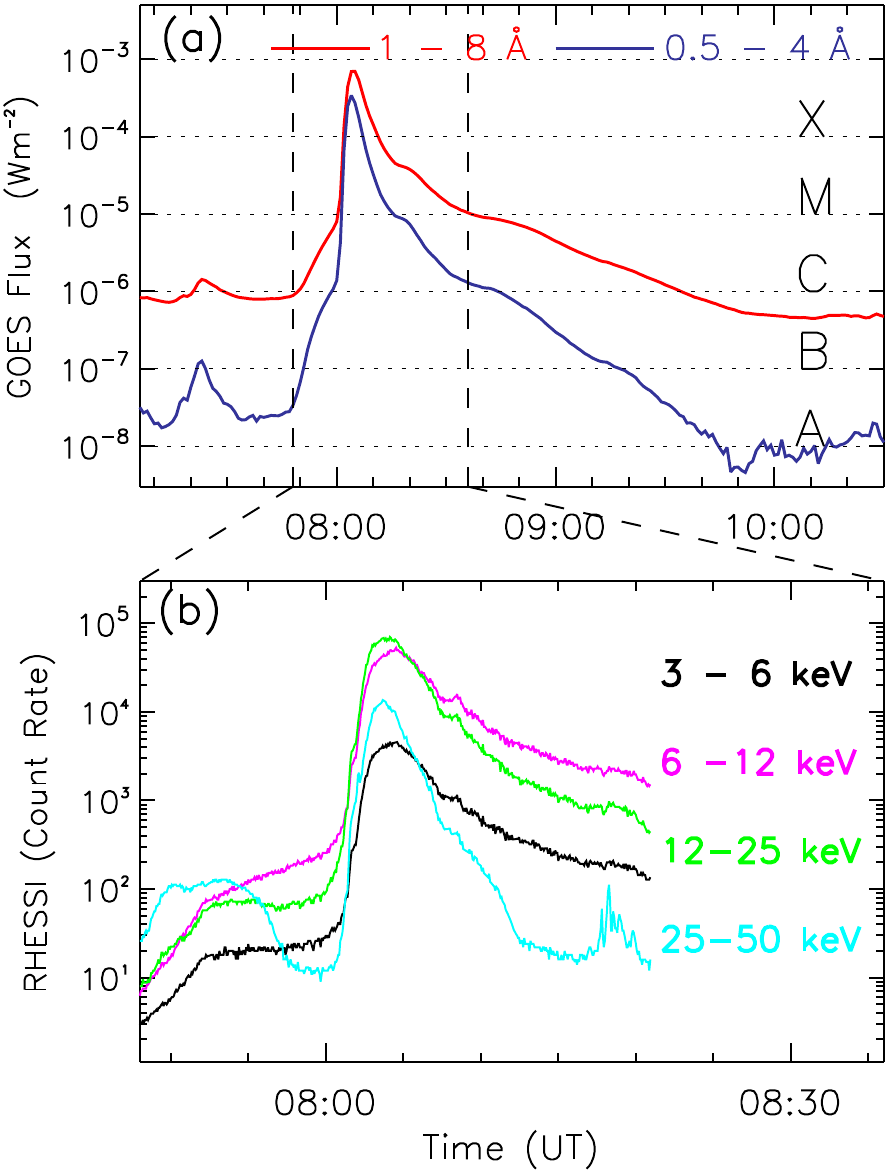}
\caption[]{\footnotesize
The soft and hard X-ray light-curves of the X6.9 flare. Top panel shows the {\em GOES} soft X-ray profiles of 0.5 -- 4 \AA\ (blue) and 1 -- 8 \AA\ (red), while the bottom one is the {\em RHESSI} hard X-ray count rates in the energy bands (4 seconds integration) of 3 -- 6 keV (black), 6 -- 12 keV (magenta), 12 -- 25 keV (green), and 25 -- 50 keV (cyan). The two vertical dashed lines in the top panel indicate the time interval of the bottom one.
\label{fig1}}
\end{figure}

\fig{fig1} shows the soft and hard X-ray light-curves of the {\em GOES} X6.9 flare occurred in NOAA Active Region AR11263 (N18W80) on August 09, 2011. The start, peak, and stop times of the flare are 07:48:00, 08:05:00, and 08:08:00 UT, respectively. During the rising phase of the flare, microwave quasi-periodic pulsations with millisecond timescale superfine structures are observed, which reflects the micro magnetic reconnection process and the dynamics of energetic electrons in the flaring process \citep{tan12}. Such microwave superfine structures are possibly important for understanding the driving mechanism of the recently discovered quasi-periodic fast propagating magnetosonic waves \citep{liu10,liu11,liu12,shen12a,shen13b,yuan13}. The flare was accompanied by a type II radio burst and a halo CME whose average speed is about \speed{1610} \citep{shen12c}. This impulsive flare also resulted in the sudden enhancement of total electron content in the sunlit hemisphere and the sudden decrease and increase of geomagnetic northward (H) component's strength at equatorial and low-latitude ionosphere of the Earth \citep{srip13}. In the low solar atmosphere, an impulsive shock wave is observed simultaneously from the top of the photosphere (say, AIA 1700 \AA) to the low corona, and it directly resulted in the oscillation of a prominence and two filaments far from the flare. According to \cite{shen12c}, the shock wave propagated with an extremely high speed of \speed{1000} at the initial stage, but quickly it decelerated to a moderate speed of \speed{605}. \cite{asai12} roughly estimated the oscillation period and amplitude of the prominence, and they are about 12 -- 16 minutes and 10 Mm, respectively. Using the Doppler and stereoscopic multi-wavelength high-resolution observations taken by the SMART, {\em SDO}, and {\em STEREO}-Ahead, in this paper we will present more details about the oscillating prominence and the other two filaments. 

\begin{figure*}[thbp]
\epsscale{0.9}
\plotone{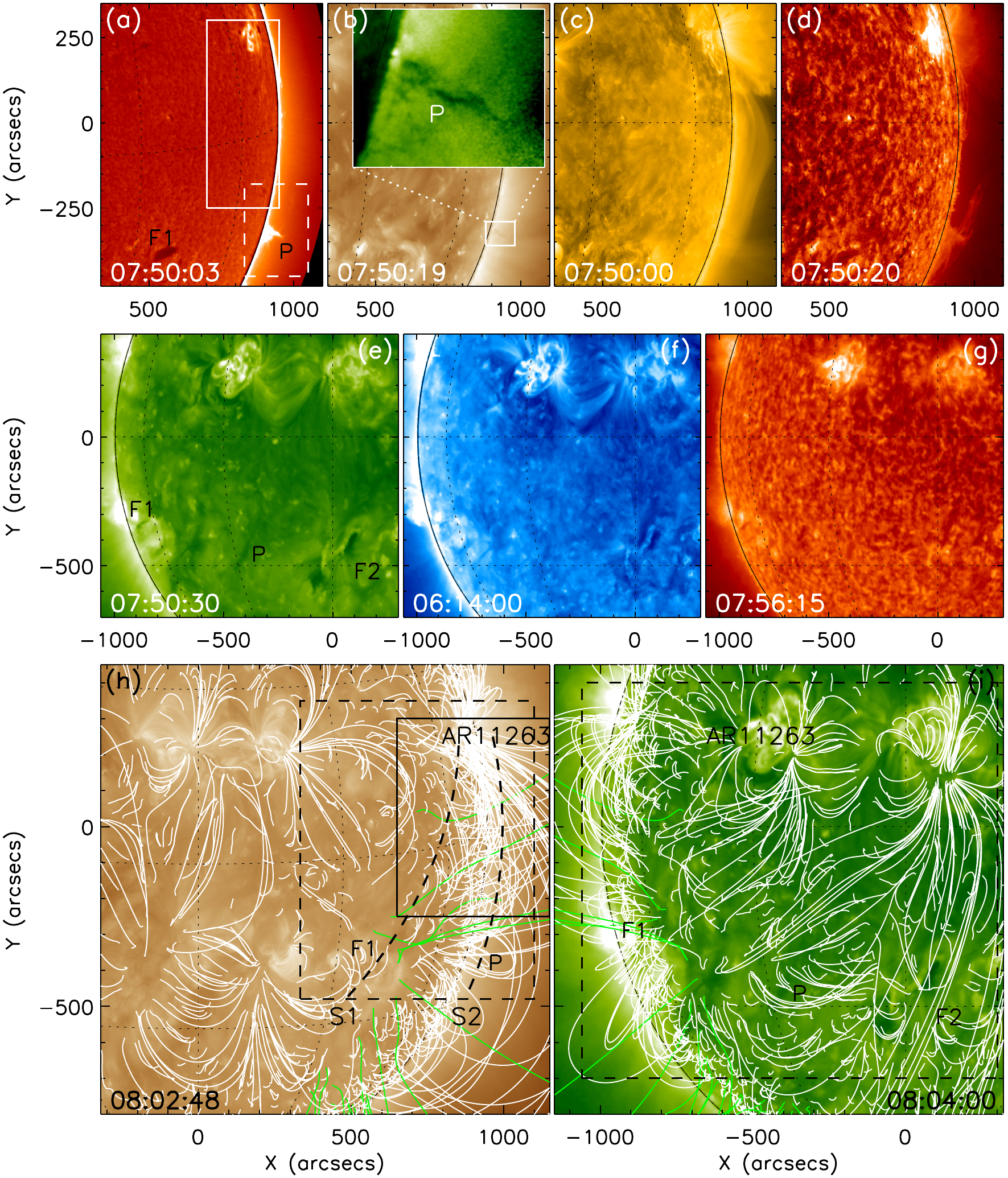}
\caption[]{\footnotesize
An overview of the event right before the flare. Panel (a) is a SMART H$\alpha$ center image (see also animation 1), panels (b) -- (d) are AIA 193 (animation 2), 171, and 304 \AA\ images, respectively. The inset in panel (b) is a closed up view of the prominence. Panel (e) -- (g) are {\em STEREO}-Ahead 195 (animation 3), 171, and 304 \AA\ images, respectively. Panels (h) and (i) are AIA 193 \AA\ and {\em STEREO}-Ahead 195 \AA\ images, respectively. They are overlaid with extrapolated three-dimensional coronal fields, in which the white (green) lines represent the close (open) field lines. The field of views of the top and middle rows are indicated by the black dashed boxes in panels (h) and (i), respectively. F1, F2, and P in panels (a) and (e) indicate the two  filaments and the prominence, respectively.
\label{fig2}}
\end{figure*}

An overview of the pre-event magnetic condition around the flare source region is shown in \fig{fig2}. From the Earth's angle of view, the location of AR11263 was close to the west limb in the northern hemisphere, while a filament (F1) and a prominence (P) can be observed in the southern hemisphere (see \fig{fig2} (a) -- (d)). From the {\em STEREO}-Ahead angle of view, AR11263 and F1 were close to the west limb, while P can be identified as a long filament along the north-south direction. In addition to F1 and P, there is another filament (F2) close to the meridian line in the {\em STEREO}-Ahead observations (see \fig{fig2} (e) -- (g)). Large amplitude oscillation of the two filaments and the prominence are observed right after the passing of the shock wave. Panels (h) and (i) are AIA 193 \AA\ and {\em STEREO}-Ahead 195 \AA\ images, respectively. They are overlaid with the three-dimensional coronal fields extrapolated with the Potential Field Source Surface (PFSS) software \citep{schr03}, where the green and white lines represent the calculated open and closed field lines, respectively. One can see that most part of the region of interest is covered with closed field except for a small region in-between F1 and the prominence, where the dominant magnetic field is open (see the green lines in \fig{fig2} (h) and (i)). With a pair of AIA 193 \AA\ and {\em STEREO}-Ahead 195 \AA\ images at 07:46:15 UT, the three-dimensional structures of F1 and the prominence can be reconstructed by using the procedure ``scc\_measure.pro'' developed by W. Thompson. The results show that the heights of F1 and the prominence are $34.8 \pm 1.5$  and $79.0 \pm 2.7$ Mm, respectively. Here the errors are given by the standard derivation of the measured data. The lower altitude of F1 could be one of the possible reason accounting for the weak oscillation of F1. The height of F2 can not be obtained because observations is only available from the {\em STEREO}-Ahead.

\begin{figure*}[thbp]
\epsscale{0.9}
\plotone{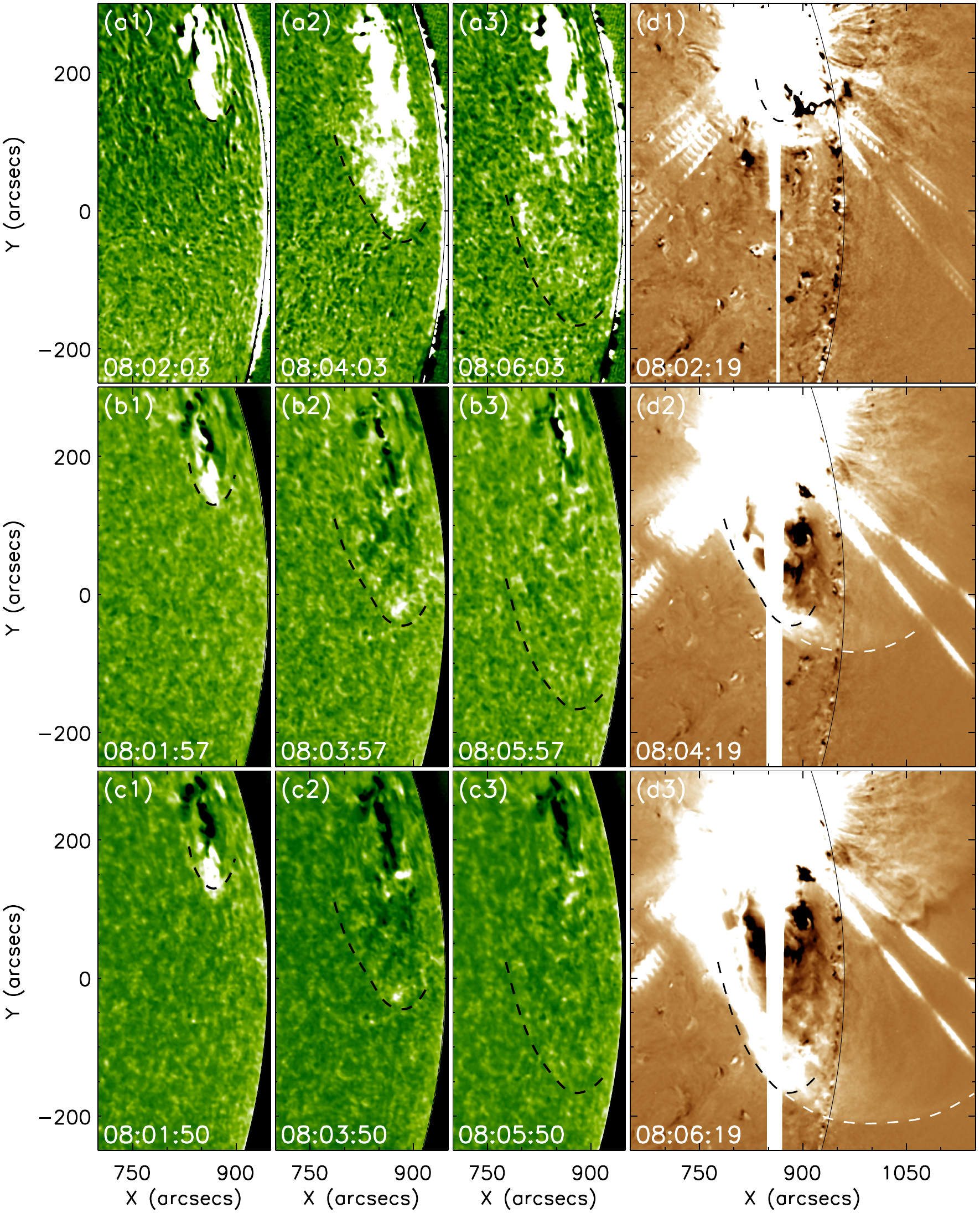}
\caption[]{\footnotesize
Chromosphere and coronal observations of the shock wave. Panels (a1) -- (a3) are SMART H$\alpha$ center based-difference images, while (b1) -- (b3) and (c1) -- (c3) are Doppler velocity images of H$\alpha$ 0.5 and 0.8 \AA\ from the H$\alpha$ line-center. (d1) -- (d3) are AIA 193 \AA\ based-difference images. The black dashed curves indicate the surface component of the shock wave, while the white dashed curves indicate the edge of the dome component of the shock wave. The FOV of the H$\alpha$ observations is shown as the white box in \fig{fig2} (a), while that for the AIA 193 \AA\ is indicated by the black box in \fig{fig2} (h).
\label{fig3}}
\end{figure*}

\begin{figure*}[thbp]
\epsscale{0.9}
\plotone{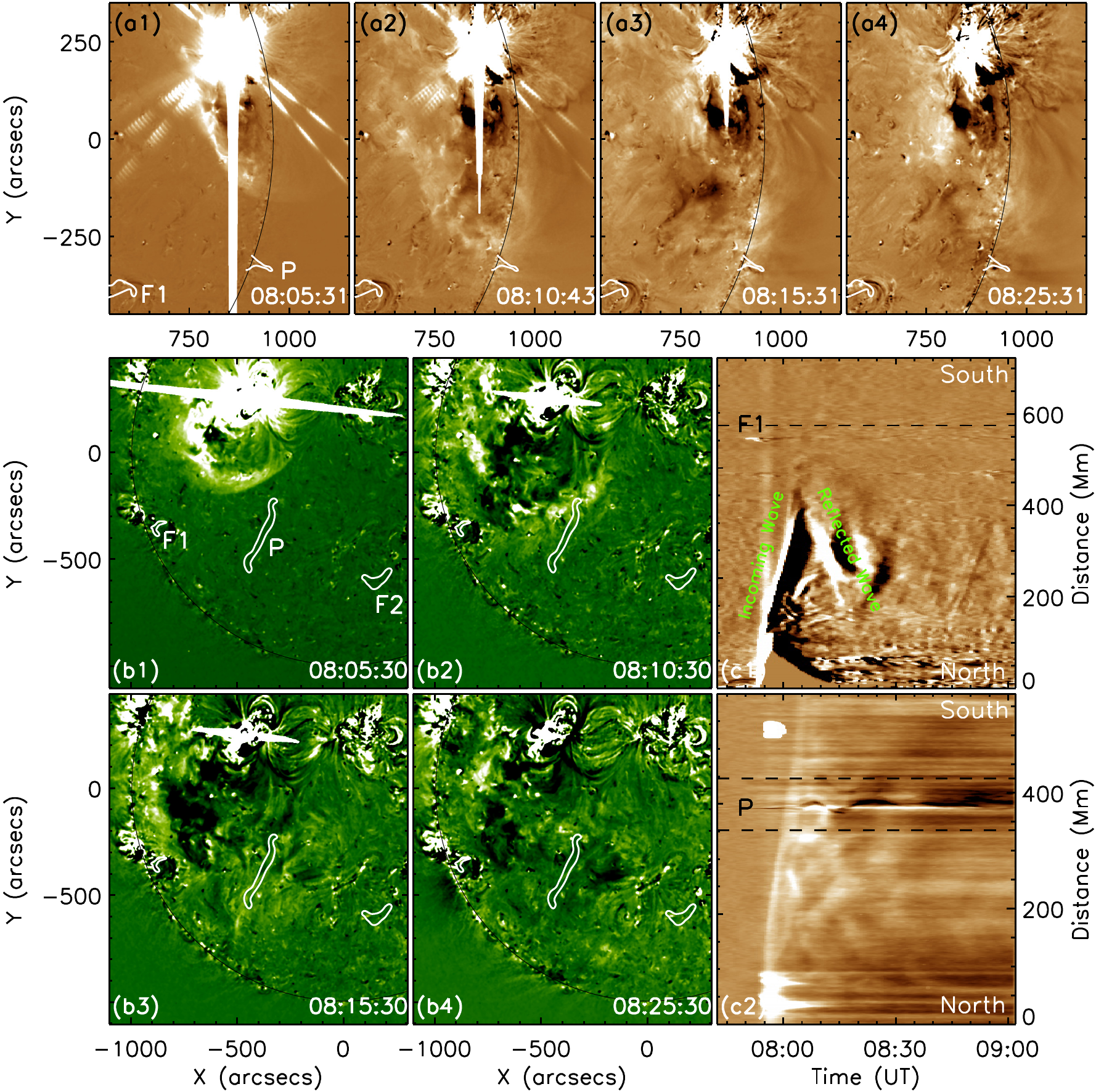}
\caption[]{\footnotesize
SDO and the {\em STEREO}-Ahead observations of the shock wave (animation 4). Panels (a1) -- (a4) and (b1) -- (b4) are SDO/AIA 193 \AA\ and {\em STEREO}-Ahead 195 \AA\ based-difference images, respectively. The white contours mark the position of the filaments and the prominence. Panels (c1) and (c2) are time-distance diagrams made from the AIA 193 \AA\ based-difference observations along S1 and S2 as shown in \fig{fig2} (h), respectively. The location of the filament (F1) and the prominence (P) are indicated by the black horizontal dashed lines in the time-distance diagrams. The field of views of (a1) -- (a4) and (b1) -- (b4) are shown as black dashed boxes in \fig{fig2} (h) and (i), respectively.
\label{fig4}}
\end{figure*}

\begin{figure*}[thbp]
\epsscale{1}
\plotone{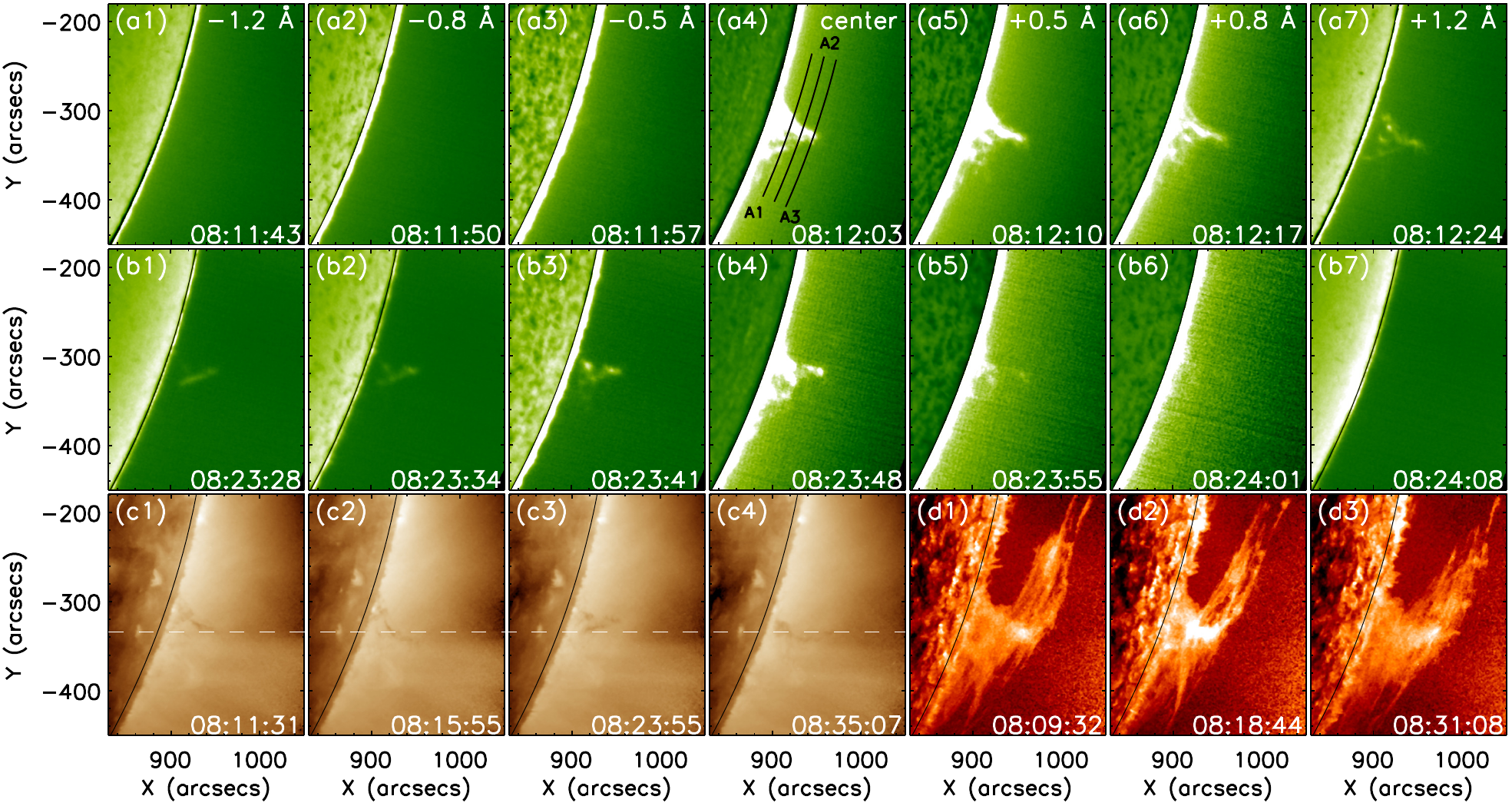}
\caption[]{\footnotesize
H$\alpha$ and EUV observations of the prominence oscillation. The top and middle rows show the SMART H$\alpha$ line-center and line-wing at about 08:12 UT and 08:23 UT, respectively. The wavelength for each column of the top two rows is indicated in the top row. Panels (c1) -- (c4) are AIA 193 \AA\ images, in which the white dashed line indicates the equilibrium position of the prominence. Panels (d1) -- (d3) are AIA 304 \AA\ observations. The field of view of each panel is shown as the white dashed box in \fig{fig2} (a).
\label{fig5}}
\end{figure*}

\begin{figure*}[thbp]
\epsscale{0.9}
\plotone{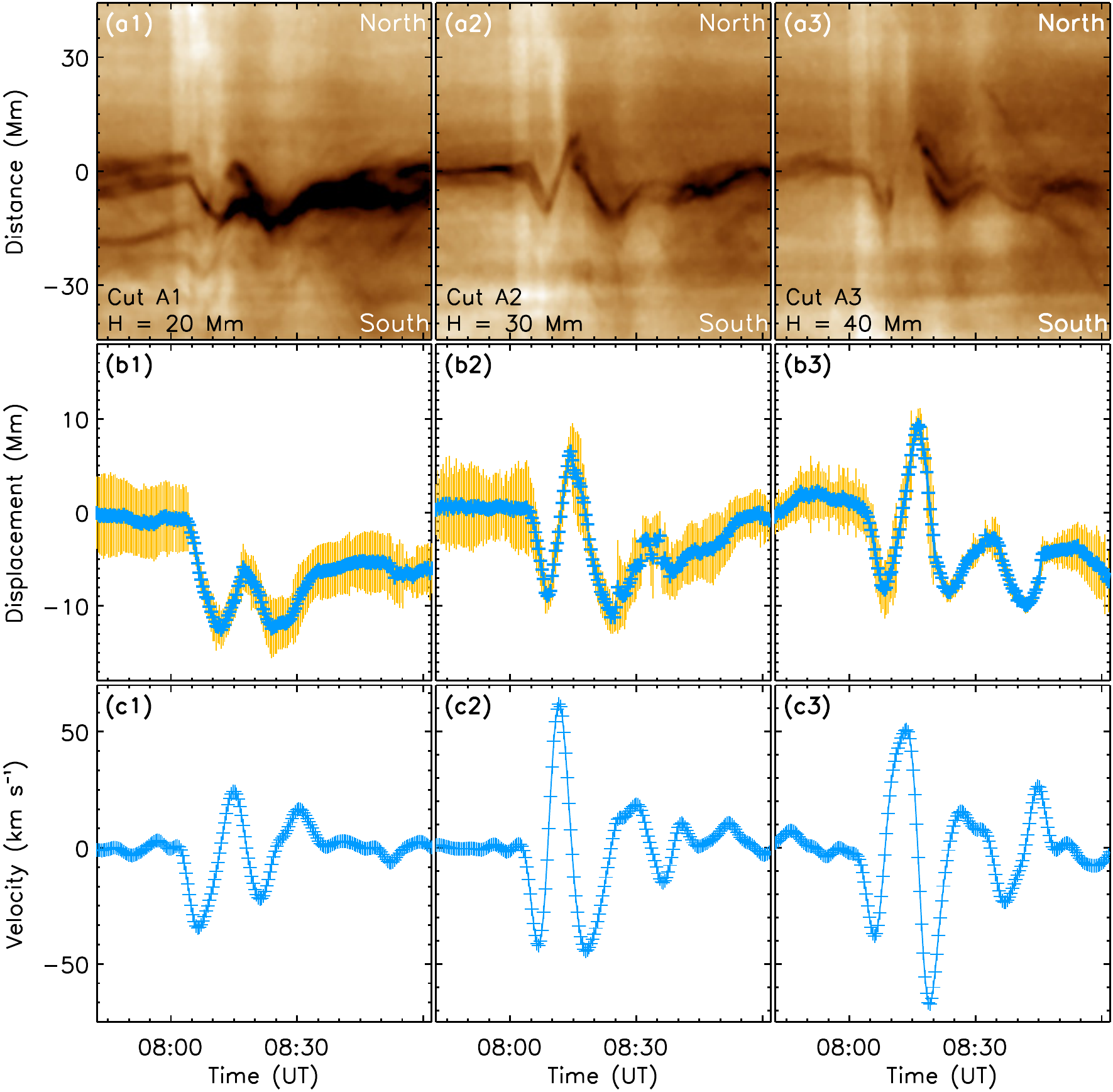}
\caption[]{\footnotesize
AIA 193 \AA\ analysis of the oscillating prominence. The top row shows the time-distance diagrams obtained from 193 \AA\ images along cuts A1 -- A3 (see \fig{fig5}(a4)), whose heights are 20, 30, and 40 Mm from the solar limb, respectively. The middle and bottom rows show the corresponding trajectory and velocity, respectively. The golden vertical bars show the measuring errors of the prominence trajectory, which are given by the widths of the Gaussian fit applied to the intensity profiles. In each frame, the solar north (south) is up (down), while the equilibrium position of the prominence is set as the original point of the prominence oscillation.
\label{fig6}}
\end{figure*}

\begin{figure*}[thbp]
\epsscale{0.9}
\plotone{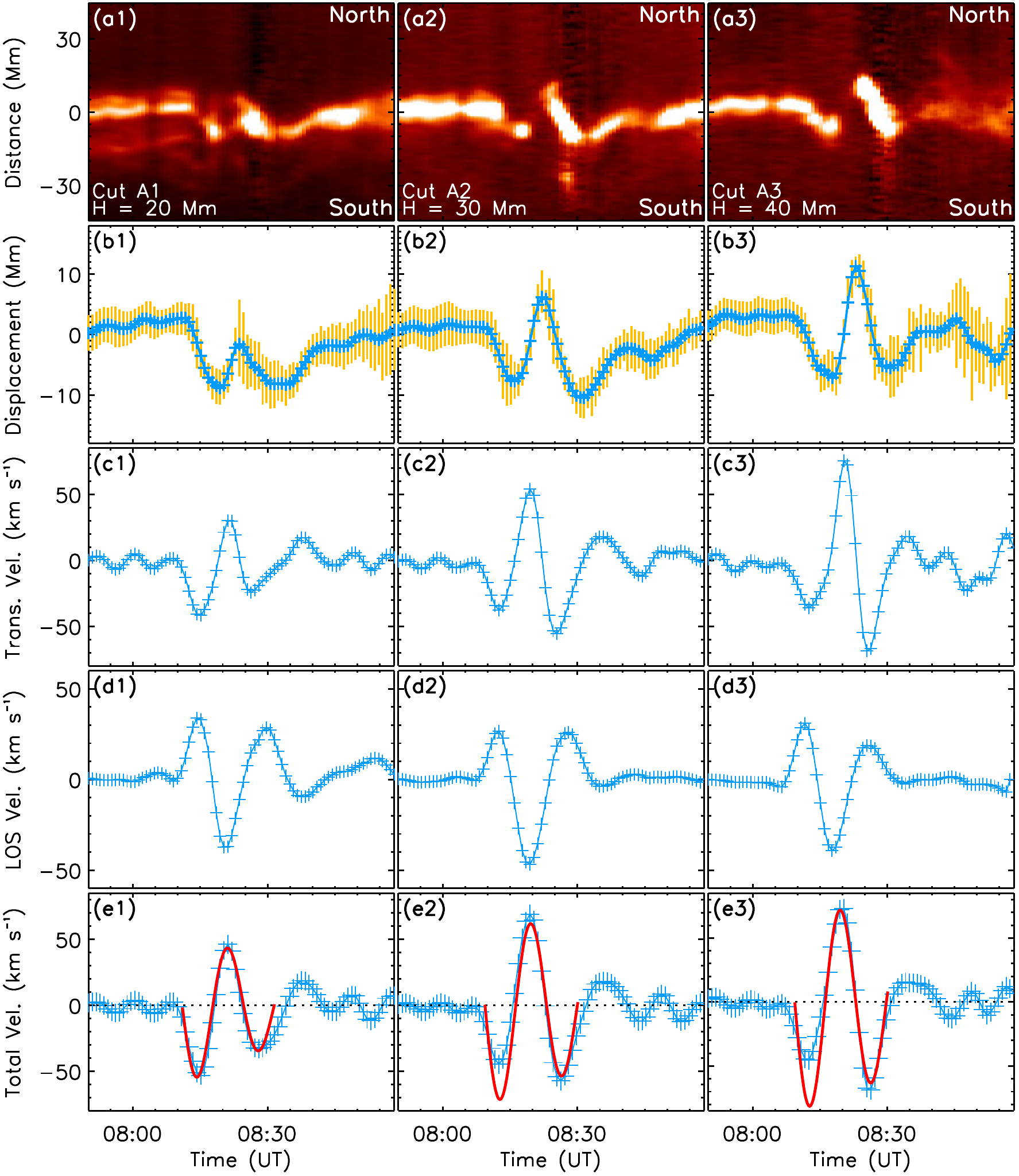}
\caption[]{\footnotesize
SMART analysis of the oscillating prominence. The top row shows the time-distance diagrams obtained from H$\alpha$ center images along cuts A1 -- A3 (see \fig{fig5}(a4)), whose heights are 20, 30, and 40 Mm from the solar limb, respectively. The second row shows the corresponding trajectories (blue) with error bars (golden) given by the widths of the Gaussian fit applied to the intensity profiles along the cuts. The third, fourth, and bottom rows show the derived transverse, LOS, and the resultant three-dimensional velocities, respectively. The red curves in the bottom row are the fitting result to the total velocity profiles. In each panel, north is up and south is down. 
\label{fig7}}
\end{figure*}

\subsection{The Shock Wave}
The shock wave was observed as a fast Moreton wave in the SMART H$\alpha$ observations. However, due to the low cadence of SMART, the Moreton wave can only be observed in three successive images (see animation 1). We show the Moreton wave in H$\alpha$ center based-difference and Doppler velocity images in \fig{fig3}. To compare Moreton wave in the chromosphere and the corresponding wave signature in the low corona, we also show the AIA 193 \AA\ based-difference images in the figure (see animation 2 and 3). It should be pointed out that the based-difference images are created by subtracting a pre-event image from the time sequence images, and the Doppler velocity images are obtained by subtracting the near-simultaneous blue-wing image from the corresponding red-wing one. In the based-difference H$\alpha$ images, the moving wavefront is seen as a white arc-shaped structure, while in the Doppler velocity images the white wavefront (red shift) is followed by a dark one (blue shift), which suggests that the dense chromosphere plasma was firstly compressed downwardly (red-shift) and then restored upwardly (blue-shift) after the passing of the shock wave. On the other hand, one can see that the coronal wave is composed of two parts in the based-difference AIA 193 \AA\ images: a sharp arc-shaped bright structure on the disk and a dome-shaped structure over the limb. By comparing the wavefronts in the chromosphere and the low corona, we find that the sharp wave structure in the corona is consistent with the chromosphere Moreton wave due to their similar structure and kinematics (see the dashed black curves in \fig{fig3} (d1) -- (d3)). As one can see from the figure, the dome-shaped structure is smoothly connecting to the sharp one. This picture indicates that the coronal shock wave should  be in a dome shape, whose lower part sweeps the solar surface and thereby forms the sharp wavefront in the corona and the Moreton wave in the chromosphere \citep{shen12c,asai12}. The EUV and H$\alpha$ observations also suggest that the wave was inclined to the solar surface. These observational results are in agreement with the theoretical interpretation in \cite{uchi68} and recent observations \citep{liu12,liu13}, where the chromosphere Moreton wave is caused by the compression of the chromosphere material by a coronal fast-mode shock wave.

\fig{fig4} shows the shock wave in AIA 193 \AA\ and {\em STEREO}-Ahead 195 \AA\ observations, in which the locations of the two filaments and the prominence are indicated by the white contours in the figure. We find that the prominence observed in AIA images is corresponding to the south leg of the filament in the {\em STEREO}-Ahead observations. It can be identified that the wave first interacted with the filament's northern leg at about 08:10:30 UT. The wave arrives at F1 and F2 at about 08:15:30 and 08:25:30 UT, respectively. When the wave approaches F1, it encounters strong reflection due to the open fields in its northern side (see \fig{fig2} (h) and (i)). The reflected wave can also be observed in the AIA observations. We show the incoming and reflected waves by means of  time-distance diagram technique, in which the propagating shock wave can be identified as a bright stripe. As one can see in the time-distance diagram, the slopes of the incoming and reflected shock wave are positive and negative, respectively (see  \fig{fig4} (c1)). One can see that the wave was reflected at a distance of about 450 Mm from the flare site, which is about 150 Mm from F1. This finding could be another possible reason for explaining the weak oscillation of F1, since the reflection of the shock wave can significantly reduce the wave energy. Since the intensity profile of coronal shock waves usually have a Gaussian shape \citep{will06}, the trajectory of the wavefront can be obtained by fitting the intensity profiles using a Gaussian model \citep[see also,][]{long11}. In addition, \cite{byrn13} found that the Savitzky-Golay filter \citep{savi64} is a more appropriate technique for deriving the kinematics of shock waves than the traditional 3-point Lagrangian method. Therefore, in the present paper we chose this method to derive the kinematics of the shock wave and the oscillating filament and prominence. The results show that the average speeds of the incoming and the reflected waves are 459.7 $\pm$ \speed{81.3} and -224.8 $\pm$ \speed{67.4}, respectively. Here the errors are the standard deviation of the derived velocities. One can see that the speed of the reflected wave is about half of the incoming wave, which indicates the significant energy loss during the interaction process. Panel (c2) shows the wave-prominence interaction process along S2, which reflects the kinematics of the dome-shaped wave component. One can see that the prominence starts to oscillate right after the passing of the shock. The average speed of the shock obtained from this time-distance diagram is 624.6 $\pm$ \speed{134.2}, which is faster than the on-disk wave component (459.7 $\pm$ \speed{81.3}). The discrepancy of the speed of the two components is possibly the result of wave delay at lower heights in the solar atmosphere due to the forward inclining structure of the shock wave \citep{liu12,liu13}, or larger magnetic field strength at lower altitude.

\subsection{The Transverse Oscillation of the Prominence}
\fig{fig5} -- 7 show the detailed analysis results of the transverse oscillating prominence. First of all, we display the H$\alpha$ Doppler and multi-wavelengths EUV observations in \fig{fig5} to show the intriguing  prominence oscillation and the wave-prominence interaction processes (see animation 4). In the H$\alpha$ observations, the prominence is located at the west limb, and it can only be observed in the H$\alpha$ center images before the oscillation. Immediately after the passing of the shock wave the prominence appeared in the H$\alpha$ red-wing observations, which indicates that the prominence was pushed to westward by the shock wave (see the top row in \fig{fig5}). Several minutes later, the prominence appeared in the blue-wing images, in the meantime, it disappeared from the red-wing observations (see the middle row in \fig{fig5}). This indicates that the prominence changed the westward motion (red-shift) to eastward (blue-shift) during this period. The motion pattern of the prominence is similar to the on-disk observations of winking filaments induced by shock waves \citep[e.g.,][]{shen14}. The observational results also indicate that the axis of the prominence is not strictly along the north-south direction, as in that case, no doppler motion could be detected in the H$\alpha$-wing observations. The oscillating prominence is also shown in AIA 193 \AA\ and 304 \AA\ images (see the bottom row in \fig{fig5}). We find that in 193 \AA\ images the prominence first moved to the south and then to the north. Taking the Doppler H$\alpha$ and the AIA imaging observations together into consideration, such a kinematics pattern suggests that the resultant motion of the prominence should be first in southwest and then in northeast direction, and the axis of the prominence is along northeast-southwest direction as shown by the {\em STEREO}-Ahead on-disk observations (see \fig{fig4} (b1) -- (b4)). In the AIA 304 \AA\ images, the interaction of the wave and the prominence can be identified by tracing the brightening caused by the compression of the wave to the prominence material (see panels (d1) -- (d3) in \fig{fig5} and animation 4), which is possibly useful for diagnosing the properties of the prominence magnetic fields and the wave energy loss of the shock wave during the interaction.

We measure the transverse motion of the oscillating prominence at three different heights, i.e., 20, 30, 40 Mm above the solar limb as indicated by the three curves in \fig{fig5} (a4). Along these cuts, we first generate the time-distance diagrams using the time sequence AIA 193 \AA\ and H$\alpha$ center images, then we are able to trace the trajectory of the oscillating prominence and to derive the oscillation velocity profile. The time-distance diagrams obtained from AIA 193 \AA\ images are shown in the top row of \fig{fig6}. The position of the prominence at a certain time is determined by fitting the intensity profile using a Gaussian profile, and the error is given by the width of the Gaussian. The traced trajectory (blue) and the associated error (golden) of the oscillating prominence is plotted in the middle row, from which we also derived the velocity profile at the three heights using the Savitzky-Golay filter method (bottom row). This figure clearly shows that the prominence starts to oscillate immediately after the arriving of the shock wave, and the oscillation lasts for about two cycles. In addition, the oscillation amplitude at higher altitude is larger than that at lower heights, which suggests that the prominence oscillate like a vertical linear rigid body with its one end anchoring on the solar surface.

\fig{fig7} shows the analysis results obtained from H$\alpha$ observations. The time-distance diagrams are shown in the top row, based on which we further derived the trajectories (second row) and transverse velocities (third row) at different heights with the same method used in \fig{fig6}. In addition, using the Doppler observations taken by the SMART telescope and the Beckers' cloud model \citep{beck64}, we also derive the Doppler velocity along the LOS. The cloud model is a widely used method for deriving parameters of cloud-like structures in the solar corona \citep[e.g.,][]{liu01a, liu01b,mori03,mori10,shen14}. To obtain the Doppler velocity of the oscillating prominence, one first needs to measure the intensity contrast profile of a point on the prominence. Here the intensity contrast is obtained by dividing the prominence intensity by the nearby background intensity. Then, fitting the intensity contrast profile with the cloud model, one can obtain the four unknown parameters in the cloud model, namely, the source function $S$, the Doppler width $\Delta \lambda_{\rm D}$, the optical thickness $\tau _{\rm 0}$, and the Doppler velocity $v$. Since the detailed description of the cloud model can be found in many articles \citep[e.g.,][and the references therein]{tzio07,shen14}, we do not describe it in the present paper. The derived Doppler velocities at different height are shown in the fourth row of \fig{fig7}. It is clear that the phase difference between the Doppler velocity and the transverse velocity is about $\pi$, which suggests that the southward (northward) motion of the prominence in imaging observations corresponds to the red-shift (blue-shift) motion in the Doppler images. This result indicates that the orientation of the prominence axis should be in northeast-southwest direction, in agreement with the on-disk observational results based on the {\it STEREO}-Ahead images. Finally, the total three-dimensional velocity can be obtained by compositing the transverse and the Doppler velocity components obtained from the H$\alpha$ observations. The results are shown in the bottom row of \fig{fig7}. To obtain the oscillation periods of the oscillating prominence at different heights, the total velocity profiles are fitted with a damping function in a form of $F(t) = A \exp(-\frac{t}{\tau}) \sin(\frac{2\pi t}{T}+\phi)$, where $A$, $T$, $\tau$, and $\phi$ are initial amplitude, period, damping time, and initial phase, respectively (see the overlaid red curves). The same fitting procedure is also applied to the transverse oscillation velocity profiles obtained from AIA 193 \AA\  (the bottom row of \fig{fig6}) and H$\alpha$ center (the third row of \fig{fig7}) observations, and the results are all listed in \tbl{tbl1}. One can see that the prominence oscillation has a larger velocity amplitude at the higher altitude, but the oscillation periods at different heights are the same (13.5 minutes). In addition, the damping time at higher height is also larger than that at lower height, which may indicates the different magnetic field and plasma density of the surrounding corona at these heights.

\begin{table*}[thbp]
\begin{center}
\caption{Various Parameters of the Oscillating Prominence. \label{tbl1}}
\begin{tabular}{ccccccc}
\tableline\tableline
\tabincell{c} {Height \\(Mm)} &\tabincell{c} {A$_{\rm Tran193}$ \\ (\speed{})} &\tabincell{c} {A$_{\rm TranH\alpha}$ \\ (\speed{})} &\tabincell{c} {A$_{\rm LOS}$ \\ (\speed{})}  &\tabincell{c} {A$_{\rm Tot}$ \\ (\speed{})} &\tabincell{c} {P$_{\rm Tot}$ \\(min)} &\tabincell{c} {$\tau_{\rm Tot}$ \\(min)}\\
\tableline
20 &33.4 $\pm$ 2.4 &37.8 $\pm$ 4.5 &33.8 $\pm$ 2.5 &65.3 $\pm$ 3.3  &13.5 $\pm$ 0.1 &31.3 $\pm$ 4.2\\
30  &52.7 $\pm$ 5.3 &55.6 $\pm$ 3.9 &38.2 $\pm$ 4.6 &78.1 $\pm$ 2.6 &13.5 $\pm$ 0.2 &45.3 $\pm$ 2.7\\
40 &84.3 $\pm$ 4.8 &82.7 $\pm$ 4.2 &40.7 $\pm$ 3.1 &89.6 $\pm$ 4.1  &13.5 $\pm$ 0.1 &54.0 $\pm$ 4.1\\
\tableline
\end{tabular}
\end{center}
\tablecomments{\footnotesize
In the table, Height is the three measured heights as shown in \fig{fig4} (a4). A$_{\rm Tran193}$ and A$_{\rm TranH\alpha}$ are the transverse velocity amplitude measured from AIA 193 \AA\ and H$\alpha$ observations, while A$_{\rm LOS}$ and A$_{\rm Tot}$ are the Doppler velocity amplitude in the LOS and the derived total velocity amplitude, respectively. P$_{\rm Tot}$, and $\tau_{\rm Tot}$ are the period and damping time obtained from the total velocity profile, respectively. The errors are the standard deviation yields by the fitting procedure.}

\end{table*}

\begin{figure*}[thbp]
\epsscale{0.9}
\plotone{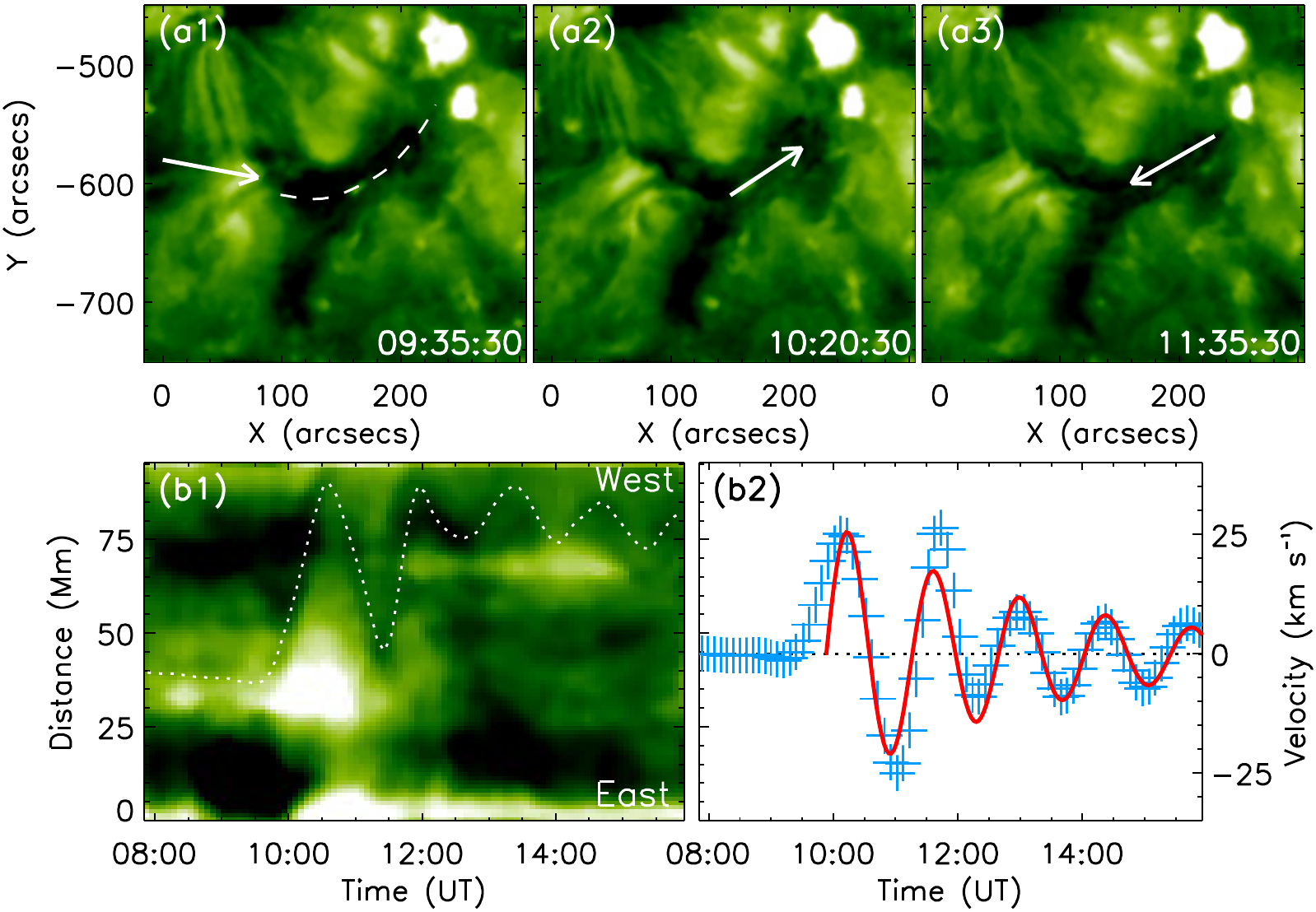}
\caption[]{\footnotesize
{\em STEREO}-Ahead observations of the longitudinal oscillation in F2 (see also animation 3). Panels (a1) -- (a3) are 195 \AA\ images, in which the white arrow in panel (a1) indicates the propagating direction of the incoming wave, while those in panels (a2) and (a3) indicate the moving direction of the oscillating filament mass. Panel (b1) is a time-distance diagram obtained from 195 \AA\ images along the filament axis as shown by the white dashed curve in panel (a1). The traced trajectory of the oscillating mass is overlaid on the time-distance diagram as a white dotted curve. Panel (b2) shows the derived oscillation velocity (blue) and the fitting result (red) of the velocity profile.
\label{fig8}}
\end{figure*}

\subsection{The Longitudinal Oscillation in F2}
From the view angle of the {\em STEREO}-Ahead, it is interesting that large amplitude longitudinal oscillation is observed in the other filament (F2) after the passing of the shock wave. In previous articles, the initiation of such kind of filament oscillation is often due to the nearby micro flares, jets, and filament eruptions. However, in the present case, it is evidenced that the longitudinal oscillation in F2 is directly caused by the shock wave. This is the first report of large amplitude filament oscillations initiated by the shock wave. Since there were no H$\alpha$ and {\em SDO} observations of this filament, we can only show the {\em STEREO}-Ahead observations in \fig{fig8}. As one can see in panels (a1) -- (a3) of \fig{fig8} and animation 3, F2 is oriented in east-west direction. The shock wave came from the east and firstly interacted with the eastern end of the filament. Then some of the filament mass started to oscillate along the main axis of the filament. The incoming wave is indicated by the white arrow in \fig{fig8} (a1), and the direction of the mass motion is indicated by the white arrows in \fig{fig8} (a2) and (a3). To study the kinematics of the oscillation, we further generate the time-distance diagram along the filament axis (dashed curve in \fig{fig8} (a1)), and the result is plotted in \fig{fig8} (b1). Since the cut is along the filament axis that is full of dark filament mass, it is difficult to trace the oscillating trajectory due to the low signal to noise ratio. Therefore, we simply trace the oscillation trajectory by eye from the time-distance diagram (white dotted curve in \fig{fig8} (b1)). Using the Savitzky-Golay filter method, the velocity profile (blue) is derived from the traced trajectory (blue curve in \fig{fig8} (b2)). The velocity profile is then fitted with the damping function $F(t) = A \exp(-\frac{t}{\tau}) \sin(\frac{2\pi t}{T}+\phi)$ (red curve in \fig{fig8} (b2)). The results show that the filament oscillated about four cycles, and the period and velocity amplitude are about 80.3 $\pm$ 3.6 minutes and 26.8 $\pm$ \speed{3.1}. It can be seen that the first cycle shows a large amplitude, and then it quickly decreases to a moderate value. This characteristic is consistent with the idea that the oscillation is launched by the arriving of the shock wave, which can exert a pulse-like pressure to the filament mass in the axial direction and thus drives the longitudinal motion in the filament.

\section{Conclusions and Discussions}
The impulsive X6.9 flare on August 09, 2011 is so far the strongest flare during the current solar cycle, and it has attracted a lot of attention of solar physicists. Several authors have studied the different aspects of this flare \citep{shen12c,asai12,tan12,sriv13,srip13}. However, a detailed study of the filament and prominence oscillations resulted from the shock wave has not been performed. In this paper, we present the first observations of simultaneous transverse oscillations of a prominence (P) and a filament (F1) and the longitudinal oscillation of another filament (F2) launched by the large-scale shock wave in association with the impulsive flare, using the high temporal and spatial resolution Doppler and stereoscopic observations taken by the ground-based SMART and space-borne {\em SDO} and {\em STEREO} instruments. In addition, we also derive the three-dimensional velocity and other parameters of the oscillating prominence, which are important for further theoretical and numerical investigations of the prominence nature.

In EUV observations, it is evident that the shock wave is composed of two distinct components: one is the sharp arc-shaped wavefront propagating on the surface ($v_{av}=$\speed{459.7}), the other is the dome-shaped wave structure over the limb ($v_{av}=$\speed{624.6}). The former has similar topology and kinematics with the wave signatures observed in AIA 1700 \AA, 1600 \AA, and H$\alpha$ observations, which image the transition region and chromosphere layers below the corona \citep[ref.,][]{shen12c}. When the sharp surface component of the shock wave encountered a region of open magnetic fields, it underwent a reflection process that reduces the wave speed to about \speed{224.8}. The speed of the reflected wave is about half of the incoming wave, which indicates the significant energy loss of the shock wave during the interaction with the open fields. The smooth connection of the two wave components suggests that the sharp wave structure on the surface should be caused by the compression of the surface plasma by the dome-shaped shock wave, in agreement with the theoretical prediction in \cite{uchi68}. The shock wave resulted in the oscillation of a prominence (P) over the limb and two filaments (F1 and F2) on the disk. Using a pair of {\em SDO} 193 \AA\ and {\em STEREO}-A 195 \AA\ images before the flare, the three-dimensional structures of F1 and the prominence are reconstructed, and their heights above the solar surface are about 34.8 and 79.0 Mm, respectively. We find that F1 showed weak oscillation after the passing of the shock wave (not shown here), unlike the strong oscillation of the prominence and F2. Based on our analysis results, we propose that the weak oscillation of F1 is possibly due to two reasons: the lower altitude of the filament and the weakening of the shock wave due to the reflection by the nearby open magnetic fields.

The prominence showed obvious transverse oscillation after the passing of the shock wave. Using the H$\alpha$ Doppler observations taken by the SMART and the Beckers' cloud model, we first derive the three-dimensional velocity of the oscillating prominence, based on the measurement of the transverse velocity in the plane of sky and the Doppler velocity along the LOS direction. The oscillation parameters at the three heights (20, 30, and 40 Mm above the surface) on the prominence are measured. The results indicate that at higher height the oscillation amplitude is larger than those at lower altitude, but the oscillation periods at different heights are all the same (13.5 minutes). This suggests that the prominence oscillate like a linear vertical rigid body with its one end anchors on the Sun. In addition, the damping time at higher height is also larger than that at lower height, which suggests that at lower heights the damping is more serious that that at higher heights. This result reflects the different magnetic field strength and plasma density of the surrounding corona at these heights. Based on the fitting results of the derived three-dimensional velocity profiles, at the heights of 20, 30, and 40 Mm above the surface the oscillation velocity amplitudes are about \speed{65.3, 78.1, and 89.6}, the periods are all 13.5 minutes, and the damping times are about 31.3, 45.3, and 54.0 minutes, respectively. F2 can only be observed from the {\em STEREO}-Ahead angle of view, therefore, we can not measure the three-dimensional parameters of this filament. However, after the passing of the shock wave, intriguing large amplitude longitudinal oscillation is observed in this filament, which is the first observation of longitudinal oscillation in filaments induced by a large-scale coronal shock wave. The longitudinal oscillation lasts for about four cycles, and it shows a large amplitude during the first cycle and then quickly decreases to a moderate value due to some unknown damping mechanisms. Our analysis results indicate that the initial velocity amplitude and oscillation period of the longitudinal oscillation are about \speed{26.8} and 80.3 minutes, respectively. We propose that the longitudinal oscillation in F2 is directly launched by the interaction of the shock wave, which can exert a pulse-like pressure to the filament mass in the axial direction and thus drives the longitudinal oscillation along the filament axis.

\begin{figure*}[thbp]
\epsscale{0.9}
\plotone{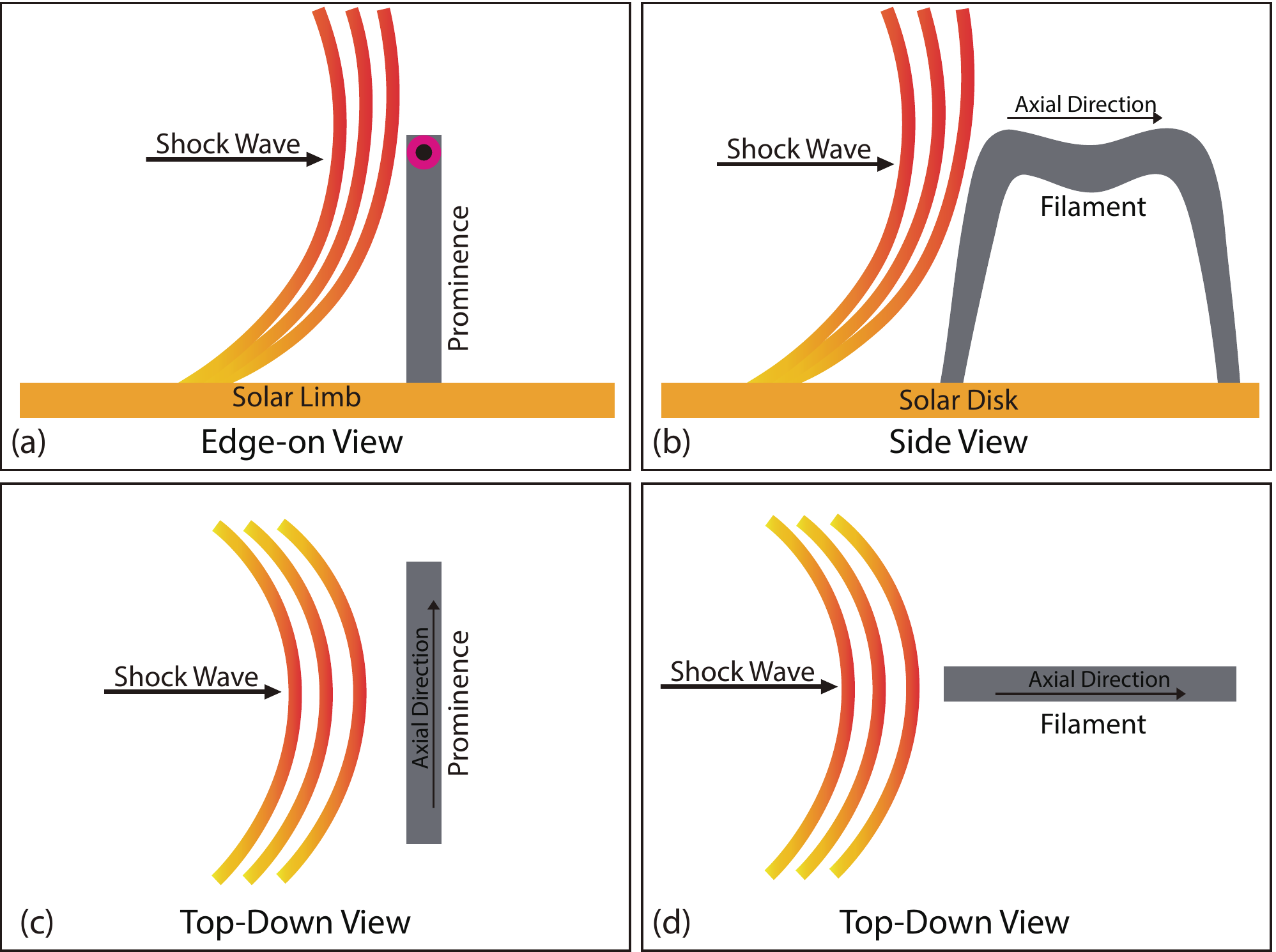}
\caption[]{\footnotesize
Cartoon demonstration of the wave-prominence and wave-filament interaction processes. Panels (a) and (b) are the edge-on and side views of the prominence and the filament, while panels (c) and (d) are the top-down view of (a) and (b), respectively. The propagation direction of the shock wave is indicated by the black arrow on the left, and red curves represent the propagating wavefront right before the interactions. The axis of the prominence is indicated by a red cycle on the top of the prominence in panel (a), while the filament axis is indicated by the thin black arrow in panel (b). The axial directions of the prominence and the filament are also indicated by the black thin arrows in panels (c) and (d), respectively.
\label{fig9}}
\end{figure*}

The stereoscopic and Doppler observation of filament oscillations is useful for diagnosing the magnetic structure and the property of filaments and shock waves, and the derived three-dimensional information of the prominence oscillation can provide important initial input parameters to further numerical and theoretical investigations of the prominence nature, because the three-dimensional results have removed the measuring errors resulting from projection effect when one uses two-dimensional imaging observations. As one can see in \fig{fig3}, the shock wave has a dome-shaped structure, and evidently the wavefront is inclined to the solar surface. Such a configuration of coronal shock waves has also been proposed theoretically \citep{uchi68} and confirmed in a few recent observational studies \citep[e.g.,][]{liu12,liu13}. Since the speed of a fast MHD wave is faster in the corona than that in the dense chromosphere, the propagation of a shock wave in the chromosphere should delay to its coronal counterpart. However, due to the low cadence of the SMART observation, it is difficult to identify such a phenomenon in this event. On the other hand, it is well-known that prominences or filaments are coronal structures that are cool plasmas supported by coronal magnetic fields \citep[e.g.,][]{kipp57,kupe74}. Therefore, the height of the interaction positions between the shock wave and the prominence and filaments should be high in the low corona. The interaction between the shock wave and the prominence presented in this paper should be a good example for demonstrating such a process. Similar to the scenario presented in \cite{vrsn02} and \cite{warm04}, we draw a cartoon in \fig{fig9} to illustrate the interaction processes between the shock wave and the prominence and the filament. We consider the prominence is a linear vertical magnetic structure with its one end roots on the Sun. When the shock wave arrives, the interaction between the wave and the prominence should take place at the top end of the prominence, and the normal vector of the shock is perpendicular to the prominence axis. The edge-on and top-down views of the wave-prominence interaction process are shown in panels (a) and (c), respectively. Due to the inclined topology of the shock wave, the pressure impacts on the prominence should point to the bottom-right direction. Therefore, the initial displacement of the prominence is naturally in this direction. After the passing of the shock wave, the prominence will oscillate freely in the corona for a few cycles. The interaction of the shock wave and F2 is shown in \fig{fig9} (b) (side view) and (d) (top-down view). Here the axis of the filament is parallel to to the normal vector of the shock wave during the interaction period. Therefore, the arriving of the shock wave will press the filament mass to move longitudinally along the filament axis. After the passing of the shock wave, the filament mass will oscillate freely in the filament channel. It should be noted that the driving mechanism of longitudinal oscillations in filament launched by a shock wave is different from those by activities of micro jets, flares, and filament eruptions, which can inject poloidal magnetic flux into the filament tube through magnetic reconnection at one of the filament legs and thereby launches the longitudinal filament oscillation \citep{vrsn07}. We would like to emphasize that a shock wave can launch not only transverse but also longitudinal filament oscillations, and the orientation of the filament relative to the wave vector should be important for launching transverse or longitudinal filament oscillation. Namely, when the normal vector of the shock wave is parallel to the filament axis, a longitudinal oscillation in the filament can be expected; otherwise, one can expect the transverse oscillation of the filament.

It is well-known that filaments or prominences are cool plasma supported by coronal magnetic field, and there are two classical models that have been widely accepted by solar physicists, i.e., the Kippenhahn-Schl\"{u}ter \citep{kipp57} and Kuperus-Raadu \citep{kupe74} models. The cool and dense filament mass resides in a magnetic dip and can keep stable due to the balance between the downward gravity and upward magnetic force. In this line of thought, the restoring force for transverse filament oscillation should be the coupling of gravity and magnetic tension. As for large amplitude longitudinal oscillations in filaments, the moving magnetized filament plasma along the filament tube would change the distribution of the filament magnetic field (especially in the case of a helical supporting magnetic field structure), which can produce a pressure gradient force opposite to the moving direction of the filament mass. As proposed by \cite{vrsn07}, magnetic pressure gradient along the filament axis can probably be the restoring force. However, if we consider that the filament mass oscillates in a curved magnetic dip, one can not ignore the gravity and magnetic tension contribution to the restoring force. Therefore, the restoring force of longitudinal filament oscillation is probably the resultant force of gravity and magnetic pressure.

Using the derived parameters of the oscillating prominence and filaments, we can estimate the strength of the supporting magnetic fields. For the transverse oscillating prominence, the radial component of the prominence magnetic field can be estimated using the method proposed by \cite{hyde66}. The relation between the radial magnetic field and the oscillation period and damping time can be written as $B_{\rm r}^{2} = \pi \rho \, r_{\rm 0}^{2} \, [4 \, \pi ^{2} \, (\frac{1}{T})^{2} + (\frac{1}{\tau})^{2}]$, where $B_{\rm r}$ is the radial magnetic component, $\rho$ is the density of the prominence mass, $r_{\rm 0}$ is the scale height of the prominence, $T$ is the oscillation period, and $\tau$ is the damping time. If we use the value $\rho = 4 \times 10^{-14} \, {\rm g  \ cm^{-3}}$, i.e., $n_{\rm e} = 2 \times 10^{10} \, {\rm cm^{-3}}$, the above equation can be rewritten as the form $B_{\rm r}^{2} = 4.8 \times 10^{-12} \, r_{0}^{2} \, [(\frac{1}{T})^{2} + 0.025 \, (\frac{1}{\tau})^{2}]$. With the measured oscillation periods and damping times, we obtain that the value of the radial component of the prominence's magnetic fields $B_{\rm r}$ at the heights of 20, 30, and 40 Mm above the solar limb are 8.13, 8.12, and 8.12 Gauss, respectively. In the calculation, we use the value $r_{\rm 0} = 3 \times 10^{9}$ cm \citep{hyde66}. The results appear to show that the prominence has the same radial magnetic field strength at different height along its main axis. However, one should keep in mind that the plasma densities at different heights are set to the same value in our calculation. For the longitudinal oscillating filament, one can derive the poloidal field in the equilibrium state with the simple equation given by \cite{vrsn07}, where the authors found that $P \approx \frac{4.4L}{v_{\rm A\varphi}}$. In the formula, $P$ is the filament oscillation period, L is an half of the filament length, $v_{\rm A\varphi} = \frac{B_{\rm \varphi 0}} {\sqrt{\mu _{\rm 0} \rho}}$ represents the Alfv\'{e}n speed based on the equilibrium poloidal field $B_{\rm \varphi 0}$. The measured L is about 54.78 Mm. With the derived oscillation period, we obtain that the Alfv\'{e}n speed $v_{\rm A\varphi} \approx $ \speed{45.48}. Using the same number density in the above calculation, we obtain that the poloidal field of the filament is about 3.22 Gauss.

It has been noted that filaments do not always oscillate when they locate on the path of a shock wave \citep[e.g.,][]{okam04,shen14}. The observation of the weak oscillation of F1 in the present case may give some hints to explain such a phenomenon. High resolution observations have revealed that shock waves always dissipate their energy to thermal or other forms energy to the ambient medium in which the wave propagates \citep[][]{thom09,shen12b,shen12c,shen13a,liuw14}. For a filament locates far from the wave's source region, the wave energy may decrease too much to launch the oscillation of the filament. In addition, as pointed out in the present paper, the low altitude of the filament and the reflection of the shock wave by the nearby magnetic structures are also important for interpreting the weak or non-oscillation filament on the wave path, since the reflection of the shock wave can significantly decrease its energy before the interaction.

In summary, the observations of large amplitude filament oscillations are important to diagnosing the properties of filaments and shock waves, as well as for prediction of a filament eruption. The restoring force and damping mechanisms of large amplitude transverse and longitudinal filament oscillations are all open questions. Especially, we present the first observations of longitudinal oscillation in a filament induced by a large-scale shock wave. Although a possible driving mechanism is proposed in the paper, it still needs more further theoretical and observational investigations in the future.

\acknowledgments {\em SDO} is a mission for NASA's Living With a Star (LWS) Program. We thank the SMART and {\em STEREO} teams for data support. The authors would also like to thank an anonymous referee for many constructive suggestions and comments, which greatly improve the quality of this paper. Y. S. is supported by the research fellowship of Kyoto University, and he would like to thank the staff and students at Kwasan and Hida Observatories for their support and helpful comments on this paper. P. F. C. is supported by the Chinese Foundations NSFC11025314 and 2011CB811402. This work is supported by the Natural Science Foundation of China (11403097), the Specialized Research Fund for State Key Laboratories, the Open Research Program of the Key Laboratory of Solar Activity of Chinese Academy of Sciences (KLSA201401), the Key Laboratory of Modern Astronomy and Astrophysics of Nanjing University, and the Western Light Youth Project of Chinese Academy of Sciences.

\end{document}